\documentclass{ifacconf}

\usepackage[round]{natbib} 
  
\usepackage{graphicx}          % Include this line if your 
\usepackage{amsmath}
\usepackage{amssymb}
\usepackage[acronym]{glossaries-extra}
\setabbreviationstyle[acronym]{long-short}
\newacronym{cdf}{CDF}{cumulative distribution function}
\newacronym{pdf}{PDF}{probability density function}
\newacronym{snr}{SNR}{signal-to-noise ratio}
\newacronym{rmse}{RMSE}{root-mean-square error}

\DeclareMathOperator{\Var}{Var}
\DeclareMathOperator{\Cov}{Cov}
\newcommand*{\cond}{\hspace*{1pt} |\hspace*{1pt}}

\begin{document}

\begin{frontmatter}

\title{Bayesian Inference for Non-Parametric Extreme Value Theory} % Title, preferably not more than 10 words.

\author{Tobias Kallehauge} 

\address{Aalborg University, Connectivity, Frederik Bajers Vej 7C, 9220, Denmark (e-mail: tkal@es.aau.dk).}

%\begin{keyword}                           % Five to ten keywords,  
%Cicero; Catiline; orations.               % chosen from the IFAC 
%\end{keyword}                             % keyword list or with the 
                                          % help of the Automatica 
                                          % keyword wizard

\begin{abstract}                          % Abstract of not more than 250 words.
Statistical inference for extreme values of random events is difficult in practice due to low sample sizes and inaccurate models for the studied rare events. If prior knowledge for extreme values is available, Bayesian statistics can be applied to reduce the sample complexity, but this requires a known probability distribution. By working with the quantiles for extremely low probabilities (in the order of $10^{-2}$ or lower) and relying on their asymptotic normality, inference can be carried out without assuming any distributions. Despite relying on asymptotic results, it is shown that a Bayesian framework that incorporates prior information can reduce the number of observations required to estimate a particular quantile to some level of accuracy. 
\end{abstract}

\end{frontmatter}

\section{Introduction}
The study of extremely rare statistical events is important to many areas of science and engineering. In meteorology and hydrology, the probability of extreme natural hazards like storms or droughts must be estimated to properly design human-built infrastructure that can withstand these dangerous events \citep{extreme_nature}. In finance, it is essential to quantify investments' risks and assess the likelihood of  negative events, e.g., stock market crashes \citep{finance}. In wireless communication, a new generation of \textit{ultra-reliable} and \textit{low-latency} networks are being deployed for critical communication applications such as automated vehicles, where characterizing the rare packet loss events is essential for safe operation \citep{wireless_access}. The fundamental problem in extreme value theory is the need to statistically quantify values of a process that are rarely or previously never observed. As such, it is necessary to extrapolate from observed to unobserved values \citep{extreme_nature}. Extreme value theory provides methods to estimate probabilities of extreme events based on limited observations, but they can suffer from model mismatch or being hyperparameter dependent. For example, the Pickands-Balkeman-de Haan theorem provides a distribution for values above some threshold, but selecting the threshold and distribution parameters can be difficult \citep{evt_wireless}. 
\\[\baselineskip]
One approach to avoid the problems of fitting distributions to rare events is to work purely with summary statistics that are estimated directly from observations without assuming an underlying distribution. Examples of such statistics include the mean and variance, which are informative statistics that can be estimated efficiently for any distribution. For extreme value theory, the quantiles for p-values with $p \approx 0$ or $p \approx 1$ are interesting statistics as they describe the ``out of distribution'' values. Summary statistics are inherently less informative than fitted distributions, which may be an issue for statistical inference. However, by choosing sufficient statistic for the application, this problem can be mitigated. Another issue is the excessive number of observations required to estimate some statistics. For example, to non-parametrically estimate the quantile for a p-value of $p = 10^{-m}$ , it is required to have at least $10^{m}$ observations \citep{marko}. 

This paper seeks to lessen the required number of observations to estimate rare-event statistics by relying on prior information. In particular, if prior information about a quantile is available, it is possible to combine this information with new observations to increase the accuracy of the estimated quantile through \textit{Bayesian inference}.

The paper is structured as follows. First, the concept of continuous quantiles and how to estimate them are introduced in section \ref{sec:evt}. The focus here will be low-value quantiles where $p \approx 0$, but the framework could be extended to studying high-values. Section \ref{sec:bay} then introduces Bayesian inference for quantiles based on prior information. Numerical results are presented in section \ref{sec:res}, where quantile estimates from the Bayesian approach are compared to a sample-based approach that does not rely on prior information. Section \ref{sec:conclusion} concludes on the general applications of paper, and section \ref{sec:perspective} gives a perspective on wireless communication as an example of where prior information can be obtained. 
% Notation
% normal distribution def, floor function, mean, variance

\section{Non-parametric extreme value theory} \label{sec:evt}
Consider the continuous random variable $X$, e.g., the duration of a drought, the share price of a company, or the \gls{snr} of a wireless communication system. A \textit{continuous quantile} with p-value $p \in [0,1]$, denoted $x_p$, is defined as the value such that the probability of $X$ being below $x_p$ is exactly $p$, i.e,
\begin{align}
P(X \leq x_p) = p.
\end{align}
For example, if $x_p = -30$ dB with $p = 10^{-5}$ in a wireless communication system, then the probability of the SNR being below $-30$ dB is $10^{-5}$. If the distribution of $X$ is known, its quantiles can be obtained through the inverse \gls{cdf}. Thus, given the \gls{cdf} $F(x) = P(X \leq x)$, the quantile is given as $x_p = F^{-1}(p)$, assuming that the inverse function exits. 

Several non-parametric quantile estimators exist, including   methods based on \textit{order statistics}, \textit{kernel density estimates}, and \textit{bootstrapping} (or some combination of these methods) \citep{quantile}. An order statistic-based method will be used here due to its tractable statistical properties, which is convenient for Bayesian inference. Thus, assume that $n$ independent observations $x^n = \{ x_i \}_{i=1}^n$ of $X$ are given. The observations are sorted in descending order such that 
\begin{align}
x_{(1)} \leq x_{(2)} \leq \dots \leq x_{(n)},
\end{align}
where $(\cdot)$ denotes the sorted index. The $l$th order statistic of $x^n$ is then simply defined as $x_{(l)}$. The \textit{sample quantile} estimate based on $x^n$ is 
\begin{align}
 \hat{x}_p = x_{(r)}, \quad r = \lfloor n p \rfloor, \label{eq:sample_est}
\end{align}
where $\lfloor \cdot \rfloor$ denotes the floor function. Intuitively, say $n = 100$ and $p = 0.2$, then the average probability that $X$ is below the $20$th order statistic is $20\%$. When either $p$ is low or $n$ is high, the case when $r = 0$ may occur. In this context, $x_{(0)}$ is undefined  since the observations do not have sufficient about the quantile. From a statistical point of view, the order statistic is unbiased and admits an asymptotic normal distribution as $n \to \infty$. This result uses the central limit theorem, and it can be shown that \citep{quantile}
\begin{align}
\hat{x}_p  - x_p \overset{d}{\to} \mathcal{N}\left(0, \frac{p(1-p)}{nf(x_p)^2}\right), \label{eq:normal_quantile}
\end{align}
where $f$ is the \gls{pdf} of $X$ and $d$ denotes convergence in distribution.  

\section{Bayesian inference} \label{sec:bay}
The result in \eqref{eq:normal_quantile} reveals that the variance of the quantile scales inversely with the number of observations $n$. To increase the accuracy of the sample quantile, a Bayesian framework can be applied when prior information is available (as mentioned, an example of this is given in section \ref{sec:perspective}). Specifically, it is assumed that the prior distribution for the quantile is normal with known parameters, i.e.,
\begin{align}
x_p \sim \mathcal{N}\left(\mu, \sigma^2 \right) \label{eq:prior}
\end{align}
Assume (for now) also that the variance of the sample quantile $\sigma_{n}^2 = \Var[\hat{x}_p] =  p(1-p)/(nf(x_p)^2)$ is known. The idea of the Bayesian framework is to combine prior information about the quantile with the new observations $x^n$. To do so, motivated by the asymptotic distribution in \eqref{eq:normal_quantile}, it is assumed that the sample quantile $\hat{x}_p$ follows a normal distribution with mean $x_p$ and variance $\sigma_{n}^2$, i.e.,
\begin{align}
\hat{x}_p \cond x_p \sim \mathcal{N}\left(x_p, \sigma_{n}^2 \right),
\end{align}
which is the \textit{likelihood distribution}. Since both the prior and likelihood distributions are normal, the posterior distribution is also a normal distribution whose parameters are available on closed form. To obtain the distribution, it is shown that
\begin{align}
E[\hat{x}_p] &= E[E[\hat{x}_p \cond x_p]] = E[x_p] = \mu \label{eq:meanderiv} \\
\Cov(\hat{x}_p,x_p) &= E[(\hat{x}_p - E[\hat{x}_p])(x_p - E[x_p])] \nonumber \\
 &= 
E[E[(\hat{x}_p - E[\hat{x}_p])(x_p - E[x_p])\cond x_p]] \nonumber \\ 
&= E[(x_p - E[x_p])(x_p - E[x_p])] \nonumber \\
&= \Var[x_p] = \sigma^2, \label{eq:covderiv}
\end{align}
\begin{align}
\Var[\hat{x}_p] &= \Var[E[\hat{x}_p \cond x_p]] + E[\Var[\hat{x}_p \cond x_p]] \nonumber \\
&= \Var[x_p] + E[\sigma_{n}^2] \nonumber \\
&= \sigma^2 + \sigma_{n}^2, \label{eq:varderiv}
\end{align}
where \eqref{eq:meanderiv} and \eqref{eq:covderiv} use the law of total expectation and \eqref{eq:varderiv} the law of total variance. % \citep{prop_theory}.
The normal posterior distribution of $x_p$ given $\hat{x}_p$ is then given by its mean 
%\citep{prop_theory}
\begin{align}
E[x_p \cond \hat{x}_p] &= E[x_p] + \frac{\Cov(x_p,\hat{x}_p)}{\Var[\hat{x}_p]}(\hat{x}_p - E[x_p]) \nonumber \\
&= \mu + \frac{\sigma^2}{\sigma^2 + \sigma_{n}^2}(\hat{x}_p - \mu) \nonumber \\
&= \frac{\sigma_{n}^2}{\sigma^2 + \sigma_{n}^2}\mu + \frac{\sigma^2}{\sigma^2 + \sigma_{n}^2}\hat{x}_p, \label{eq:bay_mean}
\end{align}
and variance 
%\citep{prop_theory}
\begin{align}
\Var[x_p \cond \hat{x}_p] &= \Var[x_p] - \frac{\Cov(x_p,\hat{x}_p)^2}{\Var[\hat{x}_p]} \nonumber \\
&= \sigma^2 - \frac{\sigma^4}{\sigma^2 + \sigma_{n}^2} \nonumber \\
&= \left(\frac{1}{\sigma^2} + \frac{1}{\sigma_{n}^2} \right)^{-1}. \label{eq:bay_var}
\end{align}
Thus, the Bayesian quantile estimate is the expectation \eqref{eq:bay_mean} with variance according to \eqref{eq:bay_var}. The expectation \eqref{eq:bay_mean} is interpreted as a weighted sum of the prior and estimated quantile with the extreme cases $E[x_p \cond \hat{x}_p] \to \mu$ for $\sigma_n^2 \to \infty$ (rely on the prior knowledge when the sample variance is high) and $E[x_p \cond \hat{x}_p] \to \hat{x}_p$ for $\sigma_n^2 \to 0$ (rely on the observations when the sample variance is low). The variance \eqref{eq:bay_var} features the inverse variances (also known as precision) and is proportional to the \textit{harmonic mean} of $\sigma^2$ and $\sigma_n^2$. Importantly; the posterior variance is less than $\sigma^2$ and $\sigma_n^2$, so the posterior estimate is generally more precise than the sample quantile.

If the variance of the sample quantile $\sigma_n^2$ is unknown, it can be estimated from the observations $x^n$. One method for this is \textit{bootstrapping}, where multiple values of $\hat{x}_p$ are estimated from random samples of $x^n$ (with replacement), and the sample variance of these then approximates $\sigma_n^2$. However, due to the simple method of estimating the sample quantile, it is possible to perform \textit{analytic bootstrapping}. Specifically, it can be shown that with infinite bootstrapping samples of size $n$, the variance estimate for the $r$th order statistic tends to
\begin{align}
\hat{\sigma}_n^2 &= \sum_{i = 1}^n \left(x_{(i)} - x_{(r)} \right)^2 w_{n,i}, \text{ with} \label{eq:bootstrap} \\
w_{n,i} &= r  \binom{n}{r} \int_{(i-1)/n}^{i/n} y^{r-1}(1-y)^{n-r} \ dy,
\end{align}
and has a relative error of the order $\mathcal{O}(n^{-1/4})$ \citep{quantile}\footnote{\citep{quantile} presents a similar estimate that has a smaller relative error, although it is not used here as it relies on a hyper-parameter that is computationally complex to estimate.}. 
\section{Numerical results} \label{sec:res}
Since the Bayesian inference relies on asymptotic results, it is not necessarily guaranteed to improve the quantile estimate particularly when $n$ is low. This section will compare the accuracy of the sample quantile in \eqref{eq:sample_est} and the Bayesian estimate in \eqref{eq:bay_mean} assuming that the hyperparameters $\mu$ and $\sigma^2$ are known.

\textit{Simulation} The quantile estimators are tested on the random variable $X = \log(Y)$, where $Y$ is exponentially distributed with rate $\lambda$. The quantile $x_p$ is simulated $1000$ times from \eqref{eq:prior} for each setting to analyze the average performance over the prior. For each simulated quantile $x_p$, $x^n$ is drawn from $X = \log(Y)$, where $F(x_p) = p$ is achieved by setting the rate of $Y$ to $\lambda = -\log(1-p)e^{-x_p}$. $x_p$ is then estimated from $x^n$ with three methods: 1) The sample quantile in \eqref{eq:sample_est}, 2) The Bayesian estimate in \eqref{eq:bay_mean} assuming $\hat{\sigma}_n^2$ is known, and 3) The Bayesian estimate where $\hat{\sigma}_n^2$ is estimated with \eqref{eq:bootstrap}. Note that the \gls{pdf} of $X$ is $f(x) = \lambda e^{x-e^x\lambda}$, which is used in the second method to compute $\sigma_n^2$ according to \eqref{eq:normal_quantile}. The \gls{rmse} between the estimates $\hat{x}_p$ and $x_p$  is then evaluated over the $10^4$ simulated quantiles. This experiment is repeated for different choices of prior variance $\sigma^2$, for different p-values $p$, and for different number of observations $n$. The prior mean is $\mu = 0$ for all settings. 
\begin{figure}
\centering
\includegraphics[width=\linewidth]{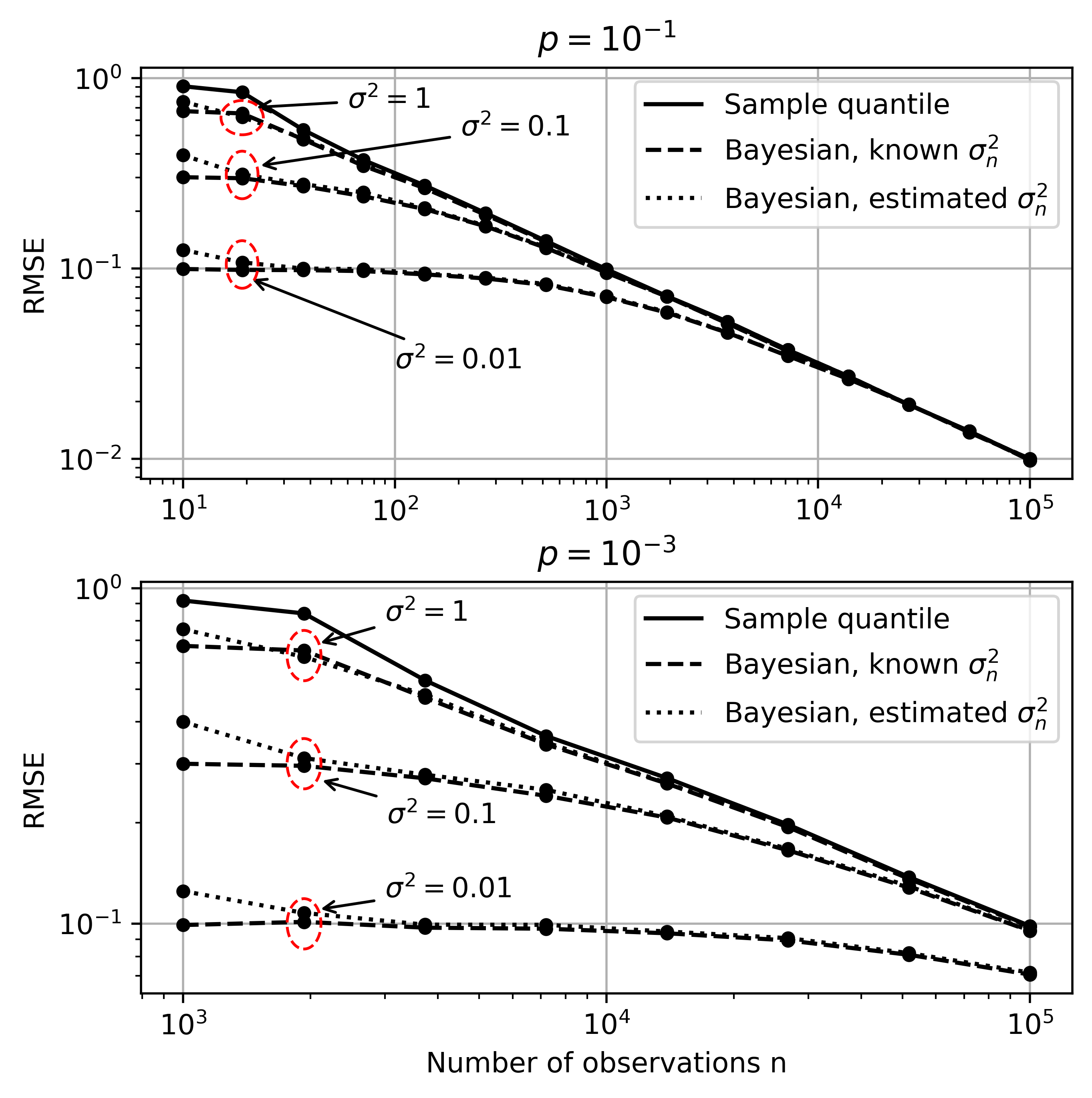}
\caption{Quantile estimation error measured by the \gls{rmse} for different p-values $p$, number of observations $n$, and prior variance $\sigma^2$. Results from the three different methods marked by solid/dashed/dotted lines and results for different values of $\sigma^2$ are denoted on the plot.} \label{fig:results}
\end{figure} 

Results from the simulations are shown in Fig. \ref{fig:results} and reveal a few interesting observations. When $n$ is low, the Bayesian approaches have lower errors, and as $n$ increases, the errors become comparable to the sample quantile approach. The prior variance is seen to significantly affect the performance of the Bayesian approaches, where the Bayesian and sample methods are similar for $\sigma^2 = 1$, and the Bayesian is about one order of magnitude better than the sample quantile for $\sigma^2 = 0.01$. Intuitively, this is because the prior information is less informative when the prior variance is high and vise versa for lower variance. Finally, it is seen that estimating $\sigma_n^2$ does not significantly impact the performance of the Bayesian approach except when $n$ is very low, where the performance slightly worsens. These observations apply to both p-values, although it is seen that the error is generally higher for lower $p$.

\section{Conclusion} \label{sec:conclusion} 
This paper discusses the problems of estimating rare-event statistics in the field of extreme value theory. To avoid the problems of parametric methods, it is proposed to work with summary statistics, e.g., quantiles. It is shown that Bayesian inference can be used to lessen the number of required samples to estimate specific quantiles for some level of accuracy. It was observed that the performance gain of the Bayesian approach depends on the prior variance. It was also observed that estimating the sample quantile variance did not significantly worsen the performance compared to known variance for the Bayesian approaches. In future works, the performance of quantile estimation for other types of distributions and the performance with uncertain prior information could be investigated.

\section{Perspectives to Wireless communication} \label{sec:perspective}
In practice, prior information about the quantiles is estimated from knowledge about the event. An example is in wireless communication with the problem of determining the \gls{snr}. Say that a mobile device at a known location wants to estimate the statistical properties of the \gls{snr}. The device can measure the \gls{snr} over a period of time. Still, it may be infeasible to acquire a sufficient number of samples  to estimate rare event statistics, such as the quantiles for low p-values. In this case, the mobile device can rely on measurements from other mobile devices in the area, i.e., prior information, to estimate \gls{snr} statistics using fewer measurements. The statistical knowledge of the \gls{snr} can then be used to select, e.g., a communication rate that ensures a reliable transmission \citep{marko}.

%\begin{ack}                               % Place acknowledgements
%Partially supported by the Roman Senate.  % here.
%\end{ack}

\bibliography{mybib}

%\appendix
%\section{A summary of Latin grammar}    % Each appendix must have a short title.
%\section{Some Latin vocabulary}         % Sections and subsections are supported  
%                                        % in the appendices.
\end{document}